\documentclass[%
 reprint,
showpacs,
nofootinbib,
amsmath,amssymb,
prc,
floatfix
]{revtex4-1}

\usepackage{graphicx}
\usepackage{dcolumn}
\usepackage{bm}
\usepackage{color}
\usepackage{hyperref}
\definecolor{LinkColor}{rgb}{0,0,0.5}
\hypersetup{colorlinks=true,%
linkcolor=LinkColor,%
citecolor=LinkColor,%
filecolor=LinkColor,%
menucolor=LinkColor,%
pagecolor=LinkColor,%
urlcolor=LinkColor}
\usepackage[caption=false, labelformat=empty]{subfig}
\usepackage{microtype}

\long\def\symbolfootnote[#1]#2{\begingroup\def\thefootnote{\fnsymbol{footnote}}\footnote[#1]{#2}\endgroup}

\begin{document}


\title{First direct measurement of resonance strengths in $^{17}$O$(\alpha, \gamma)^{21}$Ne}

\author{A. Best}
\email{abest1@nd.edu}
\author{J. G\"{o}rres}
\author{M. Couder}
\author{R. deBoer}
\author{S. Falahat}
\altaffiliation[Present address: ]{Institute for Environmental Research, Australian Nuclear Science and Technology Organisation, Kirrawee DC, Australia}
\author{A. Kontos}
\author{P. J. LeBlanc}
\author{Q. Li}
\author{S. O'Brien}
\author{K. Sonnabend}
\altaffiliation[Present address: ]{Institute for Applied Physics, Goe\-the-Uni\-ver\-si\-ty Frankfurt, Frankfurt, Germany}
\author{R. Talwar}
\author{E. Uberseder}
\author{M. Wiescher}
\affiliation{%
Department of Physics, University of Notre Dame, Notre Dame, Indiana 46556, USA 
}%

\date{\today}

\begin{abstract}
	The reaction $^{17}$O$(\alpha,\gamma)^{21}$Ne has been measured by in-beam gamma spectroscopy for the first time in the energy range $\text{E}_{\alpha} =$ 750 keV to 1650 keV using highly enriched anodized $\text{Ta}_2(^{17}\text{O})_5$ targets. Resonances were found at $\text{E}_\alpha =$ 1002 keV, 1386 keV and 1619 keV. Their strengths and primary $\gamma$-ray branchings are given. The new results exclude the low reaction rate of Descouvemont and support the rate of Caughlan and Fowler. Implications for the neutron poisoning efficiency of ${}^{16}$O in the weak s process are discussed.
\end{abstract}
\pacs{26.20.Kn, 24.30.-v, 23.20.Lv}

\maketitle

\section{INTRODUCTION}
Elements in the mass range A=60--90 are produced by neutron capture on iron seed nuclei during the core Helium and shell Carbon burning phases in massive stars.
The element production of this \emph{weak s process} depends heavily on the neutron density, the main neutron source being the reaction $^{22}$Ne$(\alpha,n)^{25}$Mg \cite{The:2000}.
During the s process, neutrons are captured by seed nuclei in the Fe region, slowly building up heavier elements.
As the rate of neutron captures is slow compared to the decay rate of unstable reaction products element production follows the valley of stability.
The neutron density in the burning environment is one of the parameters that determines the final abundance and the end point of the reaction path.
At the same time, isotopes with high abundance or high neutron capture cross sections can capture a large amount of free neutrons, thereby acting as a neutron poison in the burning environment.

Recent results from calculations of massive, fast rotating stars at low metallicity showed a large increase in the s process isotope production over the yields from non-rotating stars \cite{Pignatari:2008}.
Due to the lower abundance of heavier neutron poisons like ${}^{25}$Mg in low-metallicity conditions, $^{16}$O becomes the dominant neutron poison via the reaction $^{16}$O(n,$\gamma)^{17}$O.
The captured neutrons can then be released again by the reaction $^{17}$O$(\alpha,n)^{20}$Ne depending on the strength of the competing reaction channel $^{17}$O$(\alpha,\gamma)^{21}$Ne. 
The competition of the two reaction channels determines the influence of the poisoning effect of $^{16}$O and, thus, the mass range of the produced elements. In addition, it has a strong effect on the
expected yields \cite{Baraffe:1993, Rayet:2000, Hirschi:2008}.

The two stellar burning phases in which the s process takes place in massive stars are core Helium and shell Carbon burning at temperatures of T = 0.3 GK and 1 GK, respectively.
For the reaction $^{17}$O$+\alpha$, $\alpha$-energies of around 0.6 MeV and 1.3 MeV are therefore important. The reaction $^{17}$O$(\alpha, n)^{20}$Ne has previously been
measured in the energy range between E$_{\alpha} =$  0.6 MeV and 12.5 MeV \cite{Denker:1994,Bair:1973,Hansen:1967}, but so far no experimental data exist on the $^{17}$O$(\alpha,\gamma)^{21}$Ne
(Q = 7347.9 keV) reaction. Reaction rates given in the compilation by Caughlan and Fowler \cite{CF88} and  by Descouvemont derived from a theoretical calculation \cite{Descouvemont:1993}   
differ by up to four orders of magnitude in the relevant energy range. This leads to substantially different elemental overproductions of the s elements in low-metallicity
rotating stars \cite{Hirschi:2008}. Because of this enormous uncertainty in the reaction rate, experimental data are of critical importance as input for more
accurate s-process simulations.

\section{EXPERIMENTAL SETUP}
The $\alpha$-beam was provided by the 4MV KN accelerator at the University of Notre Dame Nuclear Science Laboratory. 
Energy calibration and resolution (1.1 keV) were determined using the well-known E$_p = 991.86 \pm 0.03$ keV and E$_p = 1317.14 \pm 0.07$ keV resonances in ${}^{27}$Al \cite{Endt:1990}.
The beam energy was reproducible within $\pm 2$ keV between different hysteresis cycles of the analyzing magnet during the course of the experiment.

The beam current on target was kept in the range of 10 $\mu$A to 30 $\mu$A in order to limit target degradation. To reduce Carbon deposition a liquid Nitrogen cooled
copper tube (cold finger) was mounted in front of the target. A bias of -400 V was applied to the cold finger for suppression of secondary electrons.
The beam was rastered with magnetic steerers to produce a beam spot of size 1.4 cm $\times$ 1.6 cm on the target. The target was mounted at $45^{\circ}$ with respect
to the beam direction and was directly water cooled with deionized water. The target chamber was electrically isolated for charge collection.

Targets were prepared by anodization of 0.3 mm thick Tantalum backings using H$_2$O enriched to 90.1\% in $^{17}$O\symbolfootnote[4]{Purchased from Isotech, Miamisburg, OH} (the $^{18}$O content of the water was specified to be 0.4\%).
This process is known to produce homogeneous films of Ta$_2$O$_5$ \cite{Vermilyea:1953,Seah:1988}.
The film thickness can be controlled in a reproducible way through regulation of the maximal anodization voltage.
The target thickness was chosen to be about 12 keV for an $\alpha$-beam of 1000 keV.

A Ge detector with an efficiency of 55\% with respect to NaI at E$_\gamma = 1333$ keV was set up at an angle of $45^{\circ}$ with respect to the beam axis.
To reduce the radiation damage from the strong $^{17}$O$(\alpha,n)^{20}$Ne channel, a polyethylene disk 8.3 cm in diameter and 2.0 cm thick was attached to
the front cap of the detector to scatter reaction neutrons away from the Ge crystal. To optimize efficiency the detector was then positioned 
in close geometry, resulting in a distance of 2.9 cm between target and detector. The target chamber and the detector itself were surrounded by at least 4.5 cm
of lead to suppress natural background radiation. The lead shielding reduced a natural background line from $^{214}$Pb at E$_\gamma = 351.9$ keV by a factor of
20, enabling a clearer signal from the dominant transition in $^{21}$Ne (E$_\gamma = 350.7$ keV). 
The detector was checked for neutron induced damage in regular intervals by monitoring the line shape of the E$_\gamma = 1173$ and 1332 keV $\gamma$-rays from a $^{60}$Co source.
No decline in resolution or change of the peak shape was observed over the course of the experiment.

\begin{figure}[htb]
	\centering
	\includegraphics[width=\columnwidth]{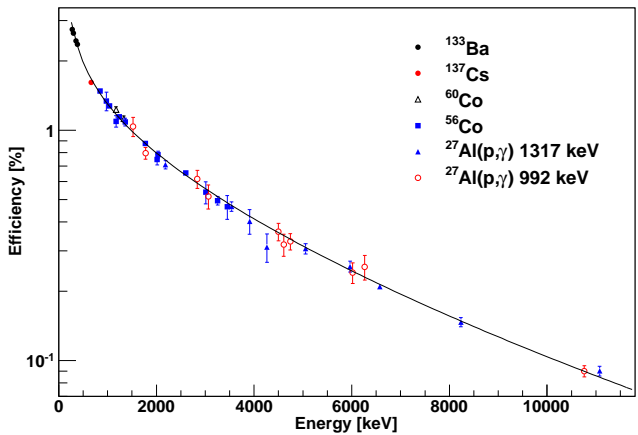}
	\caption{(Color online) Efficiency of the Ge detector as a function of photon energy. The solid line represents a fit to the data points. The data points from ${}^{56}$Co and the 1317 keV $^{27}$Al(p,$\gamma)^{28}$Si resonance are relative measurements that have been scaled to coincide with the absolute efficiencies.}
	\label{fig:eff}
\end{figure}

The absolute peak efficiency of the setup in the energy range E$_\gamma = 276$ keV to 10800 keV was determined using calibrated ($\pm$ 5\%) $^{137}$Cs, $^{60}$Co and $^{133}$Ba
sources placed onto the target as well as with the well-known strength of the E$_p= 992$ keV resonance in $^{27}$Al(p,$\gamma)^{28}$Si
($\omega\gamma = 1.93 \pm 0.13$ eV) \cite{Paine:1979}. The $^{137}$Cs source measurements were also used to determine the total efficiency
of the setup that was needed for summing corrections. The total efficiency as a function of energy was simulated using Geant4 \cite{Agostinelli:2003}. Lead shield, target chamber and holder and detector geometry were implemented as close as possible to the physical setup. The results were scaled to coincide with the
source measurement ($9.1 \pm 1.0 \%$ at E${}_{\gamma} = 662$ keV). The efficiencies from the Al resonance were corrected for summing effects using the branchings given in \cite{Endt:1990a}.

Measurements with a ${}^{133}$Ba source were performed with the detector at the close distance (2 cm) and with the detector retracted to a larger
distance (12.4 cm) to investigate summing effects on the efficiency determination. 
The influence of summing on the Ba measurements can be determined as follows: the transitions which can be affected by summing involve photons
with energies E$_{\gamma} = 81$ keV and E$_{\gamma} = 80$ keV.
There was enough absorbing material between source and detector (0.25 mm Tantalum backing, 1 mm brass target holder and the 2 cm thick polyethylene disk)
to attenuate low-energy photons by approximately 98\%, resulting in summing corrections of less than 0.5\%.
A comparison between Ba measurements in close and in far geometry showed no difference in the relative intensities of the $\gamma$-ray peaks, as would be expected without summing effects.

The Geant4 simulation was also used to investigate the difference between point and area sources on the detector efficiency. The decay radiation from a ${}^{60}$Co source was simulated, first with the source of the $\gamma$-rays in the center
of the target and then with the emission point randomly distributed on an area equal to the size of the beam spot. The effect was found to be less than 1\%.

In addition to these absolute measurements, relative efficiency data were augmented using the 1317 keV resonance in $^{27}$Al(p,$\gamma)^{28}$Si and with a $^{56}$Co source. 
These spectra were measured with the detector at the large distance and the relative efficiencies were scaled to coincide with the absolute efficiency measurements.
The resulting efficiency curve is shown together with the data in Fig. \ref{fig:eff}.

\section{EXPERIMENTAL RESULTS}
\begin{figure}[htb]
	\centering	
		\includegraphics[width=\columnwidth]{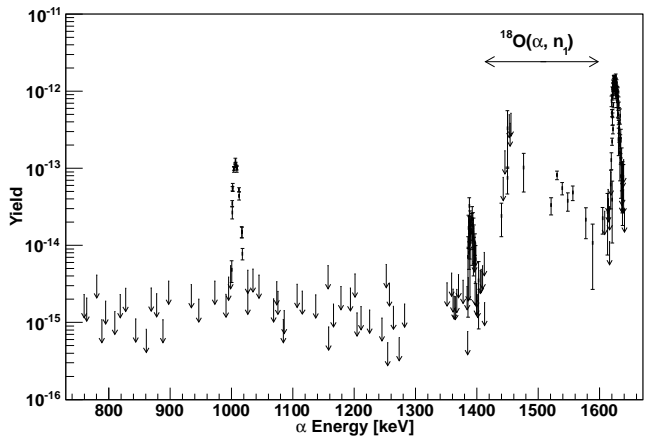}
		\caption{Yield (reactions per projectile) of the 350 keV $\to$ g.s. transition in $^{21}$Ne, not corrected for ${}^{18}$O background (see text). Room background contributions to the analysed peak have been subtracted. The high yield around 1450 keV arises from the contribution of a strong resonance in $^{18}$O($\alpha, n_1)^{21}$Ne producing the same $\gamma$-transition. The $^{18}$O($\alpha, n_1)$ yield drops significantly towards the $n_1$ threshold at E$_\alpha = 1280$ keV. The arrows denote upper limits.}
	\label{fig:yield-curve}
\end{figure}
To search for resonances in the $^{17}$O$(\alpha,\gamma)^{21}$Ne reaction, an excitation curve (see Fig. \ref{fig:yield-curve}) in the energy range of E$_{\alpha} = 750$ keV to 1650 keV was measured in steps of 10 keV or less.
For this search the E$_\gamma = 350.7$ keV transition from the first excited state to the ground state in the $^{21}$Ne compound nucleus was observed.
This transition was chosen because of the high efficiency of the detector for $\gamma$-rays with relatively low energies and because most $\gamma$-cascades proceed through the first excited state in ${}^{21}$Ne \cite{Hoffmann:1989}.
The yield $Y$ (number of reactions per projectile) was calculated from the intensity $I$ in the 350 keV peak by:
\begin{equation}
	Y = \frac{I}{Q_{dt} \eta} \; .
	\label{eq:yield}
\end{equation}
$Q_{dt} \text{ and } \eta$ represent the dead time corrected number of projectiles and the detector efficiency at 350 keV.

Because of the 0.4\% content of $^{18}$O in the targets and the strength of the $^{18}$O$(\alpha,n)^{21}$Ne reaction there can be a contribution to the counts
in the 350 keV peak from the $^{18}$O$(\alpha,n_1)$ channel which could lead to the assignment of spurious resonances in the $^{17}$O$(\alpha,\gamma)$ excitation curve.
The $^{18}$O$(\alpha,n_1)$ reaction was measured over the whole energy range covered in this experiment using the same experimental setup \cite{Best:2010a}. The contributions
due to the target contamination were subtracted from the experimental yield of the $^{17}$O$(\alpha,\gamma)$ reaction. Fig. \ref{fig:yield-curve} shows the resulting yield curve
(not corrected for the $^{18}$O contribution). The influence of the $^{18}$O$(\alpha,n_1)$ reaction is clearly visible in the energy range of 1400 keV to 1600 keV where
its strongest resonance in this energy range dominates the overall yield.

Target stability was frequently checked by scanning the E$_\alpha = 1620$ keV resonance in $^{17}$O$(\alpha,n_1)^{20}$Ne using the 1633.7 keV transition to
the $^{20}$Ne ground state \cite{Best:2010a}. When a resonance was found it was rescanned in finer steps and long runs with a fresh target were taken on top
of and just below the resonance. A total of four similar $^{17}$O targets were used over the course of the experiment.

\begin{figure}[h!]
	\centering
	\subfloat[]{\includegraphics[width=\columnwidth]{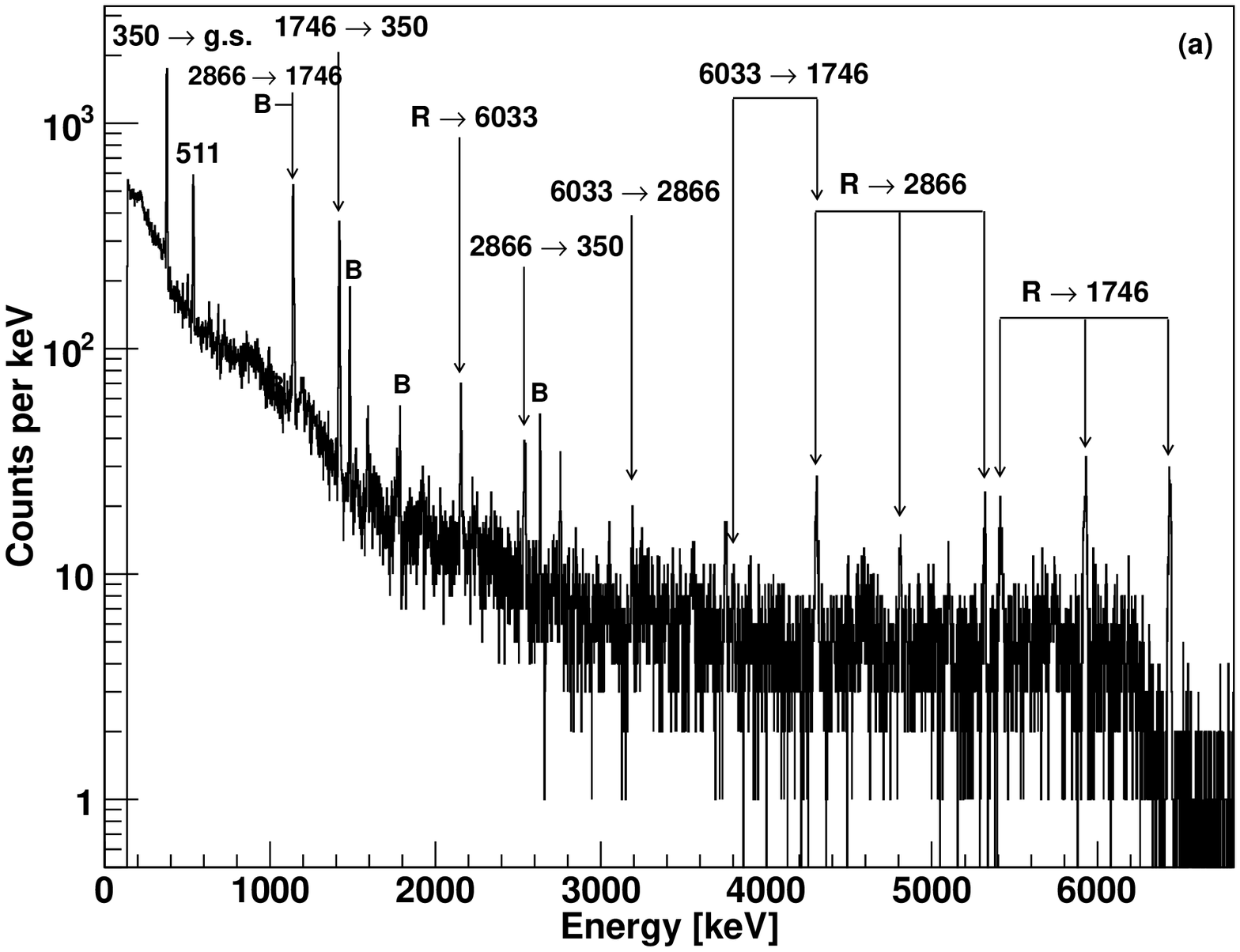}\label{fig:810}}\\
	\subfloat[]{\includegraphics[width=\columnwidth]{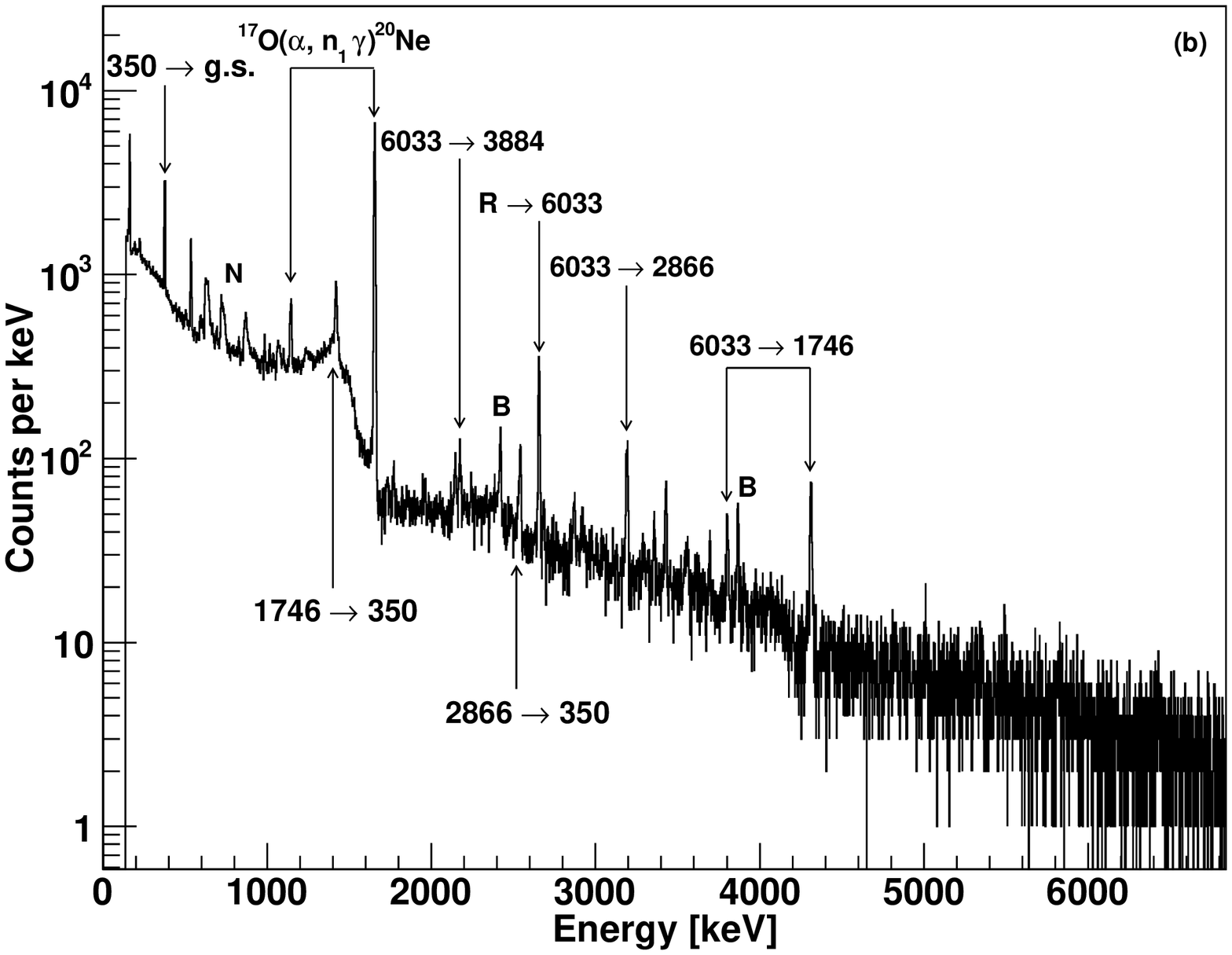}\label{fig:1311}}
	\caption{On-resonance spectra at E$_{\alpha}$ = 1002 keV (a) and 1619 keV (b). Background lines seen in off-resonance runs (B), neutron induced peaks (N) and transitions due to the decay of the resonant state are identified. In (b), the E$_\gamma = 1633.6$ keV transition from the ${}^{17}$O($\alpha, n_1){}^{20}$Ne reaction is the most prominent line.}
	\label{fig:spectra}
\end{figure}

Resonances were found at E$_{\alpha}$ = 1002 keV, 1386 keV, and 1619 keV, in good agreement with known states in $^{21}$Ne (see Tab. \ref{tab:results}).
On-resonance spectra of two of the resonances are shown in Fig. \ref{fig:spectra}. For the weak 1386 keV resonance only the 350 keV transition to the ground
state could be observed. In the $\gamma$-spectrum of the 1619 keV resonance (Fig. \ref{fig:1311}) the prominent 1633 keV line from $^{17}$O$(\alpha, n_1)^{20}$Ne
can be clearly seen as well as some peaks due to the interaction of reaction neutrons with the Germanium detector. The spectrum of the lowest energy resonance is shown in Fig. \ref{fig:810}.
The state lies below the $n_1$ threshold and the neutron yield is sufficiently low to have a much cleaner spectrum, enabling the identification of more resonant transitions.
Resonant $\gamma$ transitions were identified by comparison with off-resonance runs and compared to known transitions in ${}^{21}$Ne \cite{Endt:1990,Hoffmann:1989}.

\begin{table}[htb]
	\centering
	\caption{Resonance strengths and energies}
	\begin{ruledtabular}
	\begin{tabular}{llllll}
		\multicolumn{4}{c}{This work} & \multicolumn{2}{c}{Ref. \cite{Endt:1990}} \\
		E$_{r}^{\text{lab}}$ & E$_{\text{x}}$ & J$^{\pi}$	&	$\omega\gamma$ & E$_{\text{x}}$	& J$^{\pi}$  \\
		keV	&	keV	&									& meV 		& keV		&	\\
		\hline
		1002(2)	&	8159(2)	&	5/2 -- 11/2					&	7.6(9) &	8154(1)	&	9/2	\\
		1386(2)	&	8470(2)	&								&	1.2(2)	&	8470(10)	&	\\
		1619(2)	&	8659(2)	&	7/2 -- 11/2					&	136(17)	&	8664(1)	&	\\
	\end{tabular}
	\label{tab:results}
	\end{ruledtabular}
\end{table}

The respective resonance strengths were calculated from the yield on top of the resonances by:
\begin{equation}
	\omega\gamma = \frac{2 \epsilon}{\lambda^2} Y \; ,
	\label{eq:omegagamma}
\end{equation}
where $\epsilon$ is the effective stopping power in the center of mass system, $Y$ is the yield calculated with Eq. (\ref{eq:yield}) and 
$\lambda$ stands for the de Broglie wavelength \cite{Illiadis:2007}.
The stopping powers were calculated using the computer code SRIM \cite{Ziegler:2004}. The experimental stopping powers have an accuracy of $\pm 5\%$ for alpha particles in this energy range.
The extracted $\omega\gamma$ values, the respective resonance and level energies are given in Table \ref{tab:results}. The following contributions to the experimental uncertainty were considered: 
current integration $\pm 3\%$, absolute efficiency (dominated by the resonance strength of the 992 keV ${}^{26}$Al resonance) $\pm 7\%$, relative efficiency $\pm 5\%$,
stopping power $\pm 5\%$ and the statistical uncertainty in the number of counts in the analyzed peaks. Angular correlation effects were estimated to be less than 10\% 
for primary transitions and less than 5\% for the 350 keV line. Overall, this resulted in an experimental uncertainty for the resonance strengths of 12\% (+ the statistical error).

For the two resonant states with observed primary transitions, a possible range of J$^\pi$ assignments was deduced by comparison of the observed lower limit of $\gamma$ strengths (using $\omega\gamma = \omega \frac{\Gamma_{\alpha} \Gamma_{\gamma}}{\Gamma} < \omega \Gamma_{\gamma}$) with the recommended upper limits from \cite{Endt:1993}.
The counts in the primary peaks have been corrected for summing effects using the branchings taken from \cite{Hoffmann:1989}.

\begin{table}[htb]
	\centering
	\caption{Gamma-ray branchings of the E$_x$ = 8158 keV resonance}
	\begin{ruledtabular}
	\begin{tabular}{ccc}
		E$_x^f$ [keV]	&Intensity	&	Intensity(lit.) \cite{Hoffmann:1989}\\
		R $\to$ 1746	&	100(12)	&	100(9)	\\
		R $\to$ 2866	&	40(10)	&	39(13)	\\
		R $\to$ 6033	&	28(6)	&	12(6)	\\
	\end{tabular}
	\label{tab:branchings}
	\end{ruledtabular}
\end{table}

\emph{The E$_{\alpha} = 1002$ keV resonance:}
The laboratory resonance energy was determined from the 50\% point of the yield curve to be $1002 \pm 2$ keV corresponding to an excitation energy of E$_x = 8159 \pm 2$ keV in ${}^{21}$Ne.
Table \ref{tab:branchings} shows the observed $\gamma$ branchings of this state in comparison to previously reported values \cite{Hoffmann:1989}.
The resonance strength was determined from the yield of the secondary E$_\gamma = 350$ keV transition as $\omega\gamma = 7.6 \pm 0.9$ meV. Allowed spins of the state are in the range of 5/2 to 11/2, in agreement
with the assignment of 9/2 by \cite{Hoffmann:1989}. Considerations of the balance of the feeding and the decay of intermediate states show that about 30\% of primary
transitions are unobserved. As there are no unidentified major lines in the on-resonance spectra, these unobserved transitions should be of low relative intensity and
spread out over multiple states.

The observation of a resonance at this energy in a recent recoil separator measurement at the TRIUMF facility has been reported in a conference proceeding \cite{Taggart:2010} 
but so far no results are available for comparison.

\emph{The E$_{\alpha} = 1386$ keV resonance:}
The resonance energy was measured as $1386 \pm 2$ keV (E$_x = 8470 \pm 2$ keV).
Due to the weakness of the resonance only the prominent E$_\gamma = 350$ keV transition could be observed. The resonance strength was found to be
$\omega\gamma = 1.2 \pm 0.2$ meV. As no primary transitions could be observed and there are no branchings for this state available in the literature
it was not attempted to correct for summing effects. Thus, the resonance strength provided here should only be used as a lower limit.

\emph{The E$_{\alpha} = 1619$ keV resonance:}
The energy of the resonance is $1619 \pm 2$ keV which translates to an excitation energy in ${}^{21}$Ne of $8659 \pm 2$ keV.
As in \cite{Hoffmann:1989}, the only observed primary transition is to the 6033 keV state.
The resonance strength was determined from the intensity of the E$_\gamma = 350$ keV transition to be $\omega\gamma = 136 \pm 17$ meV. 

Due to the occurrence of resonances in ${}^{18}$O($\alpha, n_1){}^{21}$Ne as well as in ${}^{17}$O($\alpha, n_1){}^{20}$Ne at this energy \cite{Best:2010a}, the observed gamma spectrum contained
many strong background and neutron-induced lines (see Fig. \ref{fig:1311}) complicating the identification of resonant gamma-ray transitions.
Comparison of the primary transition strengths with their recommended upper limits \cite{Endt:1993} restrict the spin values of the resonant state to the range 7/2 to 11/2.

\section{DISCUSSION}
The reaction ${}^{17}$O($\alpha, \gamma){}^{21}$Ne has been measured in the energy range 750 keV to 1650 keV and three resonances have been found.
The lowest observed resonance is of comparable strength in the $\gamma$ and the neutron reaction channels ($\omega\gamma_{\gamma} = 7.6 \pm 0.9$ meV
vs. $\omega\gamma_{n} = 4.2 \pm 2$ meV \cite{Denker:1994}). This resonance is in the energy range of interest for Carbon shell burning and has a significant impact on the relevant reaction rate.

S process calculations used either the Caughlan-Fowler reaction rate for ${}^{17}$O($\alpha, \gamma)$ \cite{CF88} (CF88 hereafter) 
 or a rate based on a microscopic three-cluster model calculation by Descouvemont \cite{Descouvemont:1993}.
Since their results differ by up to four orders of magnitude in the relevant energy regions network calculations of the weak s process, especially for low metallicity stars
\cite{Hirschi:2008}, could either produce significant overabundances of elements up to mass 90 with the higher CF88 rate or extend the element production into the A = 150
region in case of a negligible neutron poisoning effect when using the reduced ${}^{17}$O($\alpha, \gamma){}^{21}$Ne rate of Ref. \cite{Descouvemont:1993}.

\begin{figure}[htb]
	\centering
		\includegraphics[width=\columnwidth]{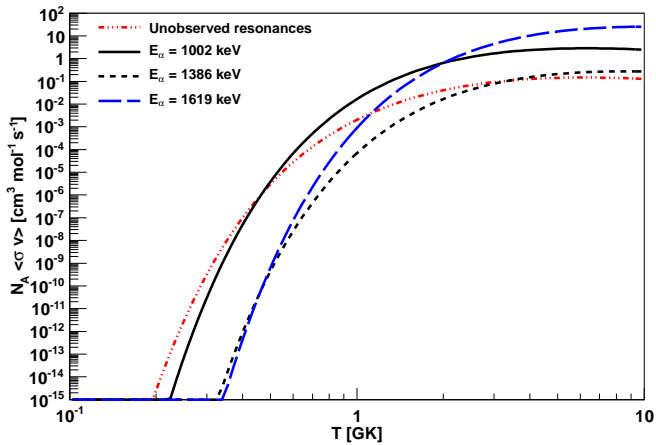}
	\caption{(Color online) Stellar reaction rate from 0.2 GK to 10 GK. The individual contributions from the three observed resonances and the sum of the unobserved states as an upper limit are shown.}
	\label{fig:rrate}
\end{figure}
Using only the three observed resonances from our experiment a calculation of the stellar reaction rate of ${}^{17}$O($\alpha, \gamma){}^{21}$Ne around a temperature of 1 GK 
confirms the CF88 estimates (at T = 1 GK: N$_{A}\langle\sigma v\rangle(present) = 1.8 \cdot 10^{-2}$ and N$_{A}\langle\sigma v\rangle(CF88) = 4.6 \cdot 10^{-2}$, in units of cm$^3$ mol$^{-1}$ s$^{-1}$) and excludes Descouvement's results.
To obtain an upper limit of the reaction rate in this energy region one can assume that all unobserved states between E$_x$= 7960 keV and 8465 keV exhibit resonances in the
$(\alpha, \gamma)$ channel with strengths just below our detection limit. This was estimated to be $10^{-15}$ reactions per projectile, corresponding to upper
limits for the strengths of possible resonances of 0.03 meV to 0.05 meV, depending on the resonance energy (see Fig. \ref{fig:yield-curve}). Their possible contribution can increase the reaction
rate at 1 GK by 10\%. Resonances at energies lower than covered in our measurement can further increase the rate. The contributions due to the individual observed
resonances and the upper limit from the unobserved ones are shown in Fig. \ref{fig:rrate}.

To cover
both burning scenarios of relevance to the weak s process it is still necessary to extend the experimental data on the reaction ${}^{17}$O($\alpha, \gamma){}^{21}$Ne
towards lower energies, if possible into the temperature range of core Helium burning around 0.3 GK. The results presented here strongly suggest that the $\gamma$-channel has
a strength that enables it to compete with the neutron channel, thereby limiting neutron recycling through the reaction ${}^{17}$O($\alpha, n){}^{20}$Ne and making
${}^{16}$O a neutron poison in the s process in massive halo-metallicity stars.
Since the critical value determining the element production is the ratio of the rates of the two competing reaction channels more detailed stellar network calculations
using the rate presented here also require the input of the ${}^{17}$O($\alpha, n){}^{20}$Ne reaction rate. We recently completed an improved measurement of this reaction
and will present an in-depth astrophysical study considering our results for both reaction channels in a forthcoming publication \cite{Best:2010a}.

\acknowledgments{The authors would like to express their gratitude to the technical staff of the Nuclear Science Laboratory at Notre Dame. This work was funded
by the National Science Foundation through grant number Phys-0758100 and the Joint Institute for Nuclear Astrophysics grant number Phys-0822648}

\bibliographystyle{apsrev}

\end{document}